\documentclass[11pt]{article}
\usepackage{moriond}

\def\be{\begin{equation}}
\def\ee{\end{equation}}
\def\bea{\begin{eqnarray}}
\def\eea{\end{eqnarray}}

\newcommand{\Photo}{}

\begin{document}
\vspace*{4cm}
\title{DISCOVERING WALKING TECHNICOLOR AT LHC AND  LATTICE
~\footnote{
To appear in Proceedings of Rencontres de Moriond EW 2013.} 
}
\author{KOICHI YAMAWAKI}

\address{Kobayashi-Maskawa Institute for the Origin of Particles and the Universe (KMI),
Nagoya University, Nagoya, 464-8602, Japan }

\maketitle\abstracts{
Walking technicolor, having a large anomalous dimension $\gamma_m \simeq 1$  and approximate scale symmetry, predicts Technidilaton,
a light composite Higgs as a pseudo-Nambu-Goldstone boson of the approximate scale symmetry,  which can be identified with the 125 GeV boson discovered at LHC. 
I will describe how such a {\it weakly coupled light composite scalar} can be dynamically realized in the {\it strongly coupled dynamics}, and can be fit to the current data at LHC, based on the ladder-like computation and the holographic one. 
I will also present results of our lattice collaboration (LatKMI Collaboration) searching for a  walking theory and a light flavor-singlet scalar bound state in large $N_f$ QCD.
}

\section{Introduction}
 A Higgs boson at  125 GeV 
has been discovered at the LHC.~\cite{Aad:2012tfa}  
Although the data so far are roughly consistent with  the standard model (SM) Higgs,
it has been reported that in the diphoton channel 
the signal strength of the boson has some tension with  the SM Higgs. 
This may imply a hint for a Higgs  boson coming from physics
beyond the SM.

\begin{figure}[h]  
\includegraphics[width=15.0cm]{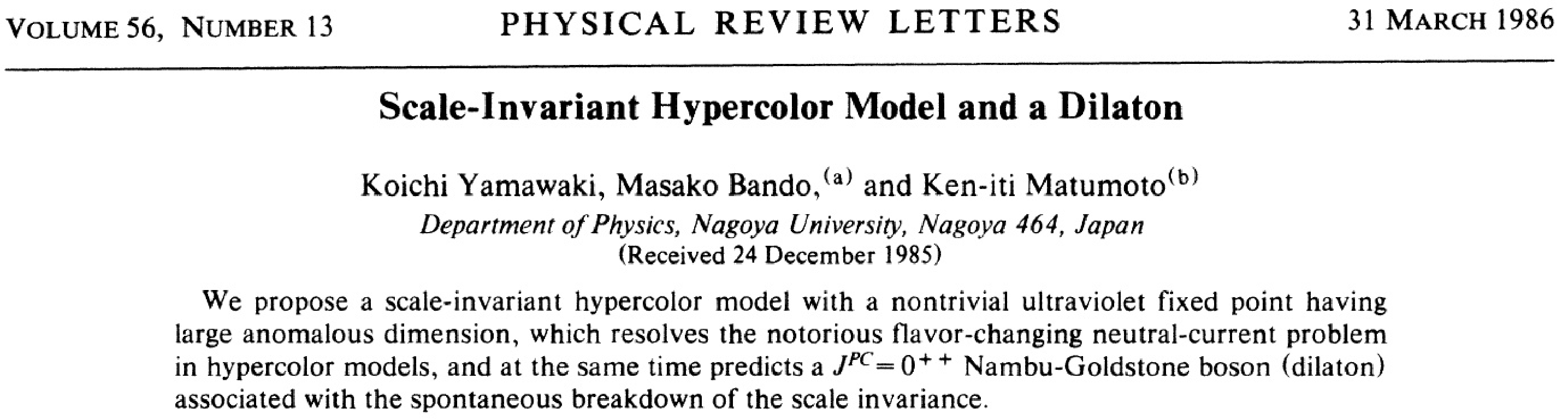}
\caption{The heading of Ref. 2.
``Technicolor'' and ``Technidilaton'' in the original manuscript were changed to ``Hypercolor'' and ``Dilaton'' , respectively,  by the  PRL Editor without consent of the authors. See INSPIRE.
 }
\label{YBM}
\end{figure}

Here we shall discuss the techni-dilaton (TD)  as a composite Higgs boson,  a technifermion and an anti-technifermion bound state,  predicted in 
the Walking Technicolor (WTC)~\cite{Yamawaki:1985zg,Bando:1986bg}  based on 
 the ladder Schwinger-Dyson (SD) equation, a concrete (approximately) scale-invariant gauge dynamics,
 which was shown to have a large anomalous dimension 
$\gamma_m =1$ as a solution
to the Flavor-Changing Neutral Currents (FCNC) problem of the original Technicolor (TC)~\cite{Weinberg:1975gm} (See Fig. \ref{YBM}).~\footnote{ 
A similar solution to the FCNC was also discussed~\cite{Holdom:1984sk}  without notion of anomalous dimension and scale invariance and hence without technidilaton.}  
TD  arises as a pseudo Nambu-Goldstone (NG) boson of the spontaneous breaking of 
the approximate scale symmetry triggered by the techni-fermion condensation which breaks the electro-weak symmetry. 
Its lightness, 125 GeV, is therefore protected  
by the approximate scale symmetry inherent to the WTC.  
 Thus the discovery of TD 
should imply discovery of the WTC.  
In fact we have demonstrated~\cite{Matsuzaki:2012mk,Matsuzaki:2012xx} 
that TD is nicely fitted to the LHC data as of July 4, 2012~\cite{Aad:2012tfa}.~\footnote{
After the Moriond EW meeting where this talk was given, new data came out, with which the TD is still consistent. See [Note added] in Ref. \cite{Matsuzaki:2013fqa}.
}

Besides a folklore against the ``lightness'' of the composite Higgs, there also exists a folklore against ``weakness of the couplings'': 
``Strongly coupled dynamics cannot lead to the composite Higgs coupled as weakly as  the 125 GeV Higgs  discovered at LHC''.
This is totally a misconception on where the strongly coupled dynamics come into play: All the bound states (techni-hadrons) including the TD are certainly strongly coupled to each other  within the WTC sector 
just as hadrons in QCD are, whereas
the couplings of 125 GeV Higgs observed at LHC are not those among the techni-hadrons but only  the couplings of the {\it techni-hadrons to the SM sector particles}, 
which must be weak, through either the (weak) $SU(2)\times U(1)$ gauge couplings or the (weak) effective
Yukawa couplings (loop-suppressed and ETC-scale suppressed via ETC-like couplings), all coming from outsides of the strongly coupled WTC sector. 
In fact we shall argue the TD couplings to the SM particles are {\it even weaker} than those of the
SM Higgs.~\cite{Matsuzaki:2012mk,Matsuzaki:2012xx}
Actually, the original TC~\cite{Weinberg:1975gm} was over-killed (three times !)  in the past, first by the problems of FCNC as mentioned and then by the S,T,U parameters, and finally by the 125 GeV Higgs at LHC.
We shall discuss that all these problems will not be applied to the WTC.

\section{Walking Technicolor}\label{sec:WTC} 
\begin{figure}[h]  
\includegraphics[width=6.0cm]{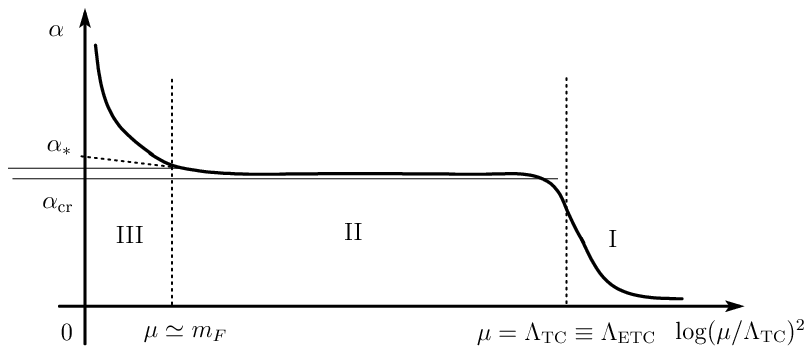}
  \includegraphics[width=6.0cm]{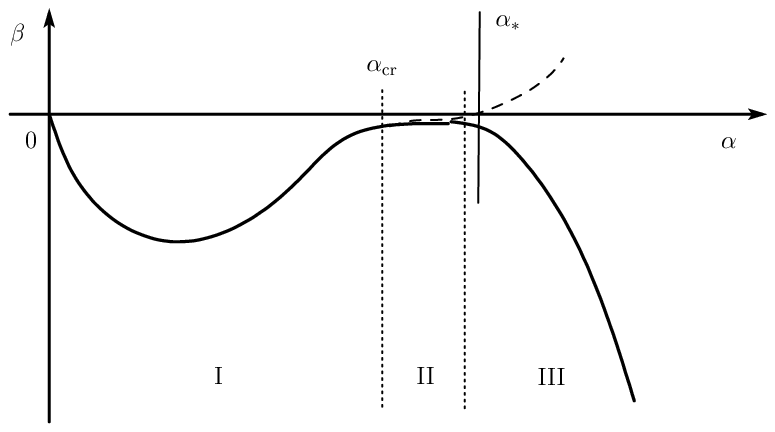}
\caption{A schematic view of WTC.  
The would-be Caswell-Banks-Zaks infrared fixed point $\alpha_*=\alpha (\mu=0)$ 
is washed out by the mass generation of $m_F$ when $\alpha_* >\alpha_{\rm cr}$. The region II corresponds to a wide region $m_F < \mu <\Lambda$ with $m_F \ll \Lambda$ in Eq. (\ref{Miranskybeta}), 
and the region III is essentially a (confining) pure Yang-Mills theory
after the decoupling of the technifermions with mass $m_F$, and the region I, $\mu >\Lambda=\Lambda_{\rm TC} $, is the asymptotically free region similar to QCD asymptotics, 
where  
the intrinsic scale 
$\Lambda_{\rm TC}$, an analogue of $\Lambda_{\rm QCD}$, is typically taken as the ETC scale $\Lambda_{\rm ETC}$. } 
\label{alpha-beta}
\end{figure}

Let us begin with briefly summarizing the WTC given in Refs.~\cite{Yamawaki:1985zg,Bando:1986bg} :
In the ladder SD equation with strong coupling $\alpha > \alpha_{\rm cr} (= \frac{\pi}{3 C_F})$, there exists a spontaneous-chiral-symmetry breaking solution with 
the mass function of the fermion $\Sigma(Q)$ ($Q^2=-q^2>0$),
$\Sigma (Q) \sim m_F^2/Q \quad (\Lambda \gg Q
 \gg m_F)
 $, which is compared with the Operator Product Expansion to yield  $\gamma_m =1$, where the dynamical mass $m_F$  ($\Sigma(m_F)=m_F$) is given by
the form of essential singularity (so-called Miransky scaling)~\cite{
Miransky:1984ef}:
\begin{eqnarray}
 m_F 
 \simeq 4 \Lambda\cdot
 \exp \left(-\frac{\pi}{\sqrt{\frac{\alpha}{\alpha_{\rm cr}} -1}}\right) \ll \Lambda \quad (\alpha \simeq \alpha_{\rm cr})
 \,,
\label{Miransky}
\end{eqnarray} with $\Lambda$ being  the cutoff introduced to the SD equation  (usually identified with 
the Extended TC scale $\Lambda_{\rm ETC}$) 
and $C_F$ the quadratic Casimir of the fermion of 
the fundamental representation of the gauge group.  
Such a dynamical generation of 
$m_F$ by the nonperturbative dynamics  in Eq.(\ref{Miransky}) should imply 
the {\it nonperturbative running}  of the coupling  
$\alpha=\alpha(\mu)$,  
even when it is perturbatively nonrunning (conformal):
\begin{eqnarray}
\beta_{_{\rm NP}} (\alpha) =
\Lambda \frac{\partial \alpha}{\partial\Lambda} = 
-\frac{2\alpha_{\rm cr}}{\pi} 
 \left(\frac{\alpha}{\alpha_{\rm cr}}-1\right )^{\frac{3}{2}} \ll 1\,,
 \quad \alpha(\mu) = \alpha_{\rm cr}\left(1 + \frac{
 \pi^2 
 }{
\ln^2\left( 
 4\mu
 /
 m_F 
 \right)}
 \right)   
 \approx \alpha_{\rm cr} \, ,
\label{Miranskybeta}
\end{eqnarray}
for $\mu\gg m_F$,  where the coupling $\alpha(\mu)$ is now slowly running (``walking'') down to $\alpha_{\rm cr}$ which is identified as  the ultraviolet fixed point~\cite{Miransky:1984ef}.~\footnote{
The beta function in Eq.(\ref{Miranskybeta}) 
has a multiple zero but not a linear zero, which implies that the beta function should turn over to the region $\alpha<\alpha_{\rm cr}$ in such a way that   
$\alpha_{\rm cr}$ is viewed as an infrared fixed point from the side of this region.}    
( A schematic view of the WTC coupling 
 in the light of Caswell-Banks-Zaks infrared fixed point~\cite{Caswell:1974gg}
 is given in Fig. \ref{alpha-beta}.)
 Then the scale symmetry is {\it broken explicitly as well as spontaneously},
by the nonperturbative running induced by the dynamical generation of  
$m_F$,
the  same origin as the spontaneous breaking,  which leads to the scale anomaly:
 $\langle \partial_\mu D^\mu\rangle=  \langle \theta^\mu_\mu \rangle 
  = -\frac{\beta_{\rm NP}(\alpha)}{4 \alpha} \langle G_{\mu \nu}^2 \rangle
 =-{\cal O} (m_F^4) \ne 0
  $,
where $D_\mu$ and $\theta_\mu^\mu=\partial_\mu D^\mu$ are the dilatation current and the trace of the energy-momentum tensor $\theta_{\mu\nu}$, respectively, of  the WTC sector,
  and $\langle G_{\mu\nu}^2 \rangle$ is the techni-gluon condensate (with the perturbative contributions subtracted), 
with $\beta_{\rm NP}(\alpha)$ being the beta function given in Eq.(\ref{Miranskybeta}).
This yields a non-zero mass of TD as a pseudo NG boson, which may be estimated through 
the Partially Conserved Dilatation Current (PCDC) relation:
 \begin{eqnarray} 
  F_\phi^2 M_\phi^2    =  -d_\theta \langle \theta^\mu_\mu \rangle 
  = \frac{\beta_{\rm NP}(\alpha)}{\alpha} \langle G_{\mu \nu}^2 \rangle
 \simeq  \kappa_V \left(\frac{2 N_{\rm TC} N_{TF}}{\pi^2}
 \right) m_F^4, 
 \label{PCDC}
  \end{eqnarray}
 where $d_\theta$ $(=4)$ is the dimension of  $\theta_{\mu\nu}$, and $M_\phi$ and $F_\phi$ are respectively the mass and the decay constant of TD denoted by $\phi$, and
 an explicit ladder computation reads $\kappa_V \simeq  0.7$
 ~\cite{Hashimoto:2010nw}.

 \section{Discovering Walking Technicolor at LHC}
  \label{subsec:TDpheno}
Now we come to the TD phenomenology at LHC~\cite{Matsuzaki:2012gd,Matsuzaki:2012mk,Matsuzaki:2012xx}.    
For concreteness we here discuss a specific model, one-family model (1FM)~\cite{Farhi:1980xs},  with 4-weak-doublets $N_D=4$ %of 
(colored techniquark doublet and uncolored technilepton doublet),  and hence the number of flavors $N_{\rm TF}=2N_D= 8$ in the $SU(N_{\rm TC})$ gauge theory.\footnote{ 
As a model-building point of view,  we may generalize the
1FM by freely adding $N_{\rm TF}^{singlet}$ weak-singlet technifermions, if necessary to satisfy the walking behavior, in which case
 $N_{\rm TF}=2 N_D + N_{\rm TF}^{singlet} \geq 2N_D $. 
 The ladder estimate of $\alpha_{\rm cr}$ combined with the two-loop value of
the Caswell-Banks-Zaks infrared fixed point $\alpha_*$ suggests~\cite{Appelquist:1996dq}  that the walking theory with $SU(N_{\rm TC})$  is close to the criticality $\alpha_*\simeq \alpha_{\rm cr}$,
or equivalently, the number of technifermion flavors $N_{\rm TF}$ should be just below $N_{\rm TF}^{\rm cr}\simeq 4 N_{\rm TC}$.
For lattice results, see section \ref{sec:Lattice}.
 } 
  We have seen 
that the simplest model, the one-doublet model (1DM), having just one weak-doublet,
generally yields too small couplings and is invisible at LHC (and hence ruled out as a TD candidate for the 125 GeV Higgs) in our framework.

Let us begin with the effective theory for the WTC which should be written in terms of the technipions $\pi$ and TD $\phi$  as the nonlinear realization of 
the scale symmetry  (up to scale anomaly) as well as the chiral symmetry, with the same results being obtained from the Ward-Takahashi identities.~\cite{Matsuzaki:2012gd}
 The chiral/EW- and scale-invariant Lagrangian thus takes the form 
 \begin{equation}
{\cal L}_{\rm inv} = \frac{v_{\rm EW}^2}{4} \chi^2 {\rm tr}[D_\mu U^\dag D^\mu U] + {\cal L}_{\rm kin}(\chi) 
\,, \label{inv:L}
\end{equation}
where $\chi(x)=e^{\phi(x)/F_\phi}$ is the field transforming linearly under the scale symmetry with the scale dimension 1 and $U(x)=e^{2 i \pi(x)/v_{\rm EW}}$ ($v_{\rm EW}=246$GeV) the chiral field expressed by  
the (partly eaten) Nambu-Goldstone boson fields $\pi$, with  $D_\mu U= \partial_\mu U - i W_\mu U + i U B_\mu$  
being a covariant derivative by 
the   $SU(2)_W \times U(1)_Y$ gauge fields 
$W$ and $B$ (disregarding $SU(3)_c$ gauging for the moment), 
 and ${\cal L}_{\rm kin}(\chi)$ denotes the scale invariant kinetic term of $\phi$. (It is straightforward to introduce the techni-$\rho/a_1$ mesons
into this Lagrangian via Hidden Local Symmetries~\cite{Bando:1987br}, implying rich phenomenology at future LHC.) 
  From the $|D_\mu U|^2$  term we read off 
the TD couplings to the massive weak bosons: 
$ 
g_{\phi WW/ZZ} = \frac{2 m_{W/Z}^2}{F_\phi}. 
$ 

The scale (as well as chiral) symmetry of the WTC sector is explicitly broken by the ETC (responsible for the SM fermion masses) 
and the  
SM gauge interactions, (a part of) which may be introduced into a scale-invariant manner using a spurion field $S(x)$ 
transforming in the same way as $\chi$, with $\langle S\rangle=1$ breaking the symmetry:
\begin{eqnarray} 
{\cal L}_{S} 
= 
- m_f \left( \left( \frac{\chi}{S} \right)^{2-\gamma_m} \cdot \chi \right) \bar{f} f 
+  \log \left( \frac{\chi}{S} \right) \left\{ 
\frac{\beta_F(g_s)}{2g_s} G_{\mu\nu}^2 
+ 
\frac{\beta_F(e)}{2e} F_{\mu\nu}^2 
\right\}  
\,, \label{Lag:S}
\end{eqnarray}
where $G_{\mu\nu}$ and $F_{\mu\nu}$ respectively denote the field strengths for QCD gluon and photon fields; 
$g_s$ and $e$ are the QCD gauge and electromagnetic couplings, respectively; 
$\beta_F$s are  the beta functions only including the technifermion 
loop contributions.  Here we disregard the explicit breakings of the chiral symmetry (coming from $|D_\mu U|^2$  term as well as ${\cal L}_{S}$) which are responsible for the mass of technipions
(See Ref.~\cite{Jia:2012kd} for details).

The TD Yukawa coupling to the SM $f$-fermion reads from the first term of ${\cal L}_S$ in Eq.(\ref{Lag:S}) 
as~\cite{Bando:1986bg}, 
$ g_{\phi ff} =  \frac{(3-\gamma_m) m_f}{F_\phi} ,
$ 
along with scale dimension of technifermion bilinear operator, 
$(3-\gamma_m)$. 
We may take $1<\gamma_m<2$ for the third-generation 
SM $f$-fermions like $t, b, \tau$, as in the 
the strong ETC model~\cite{Miransky:1988gk} where the condensate takes place by the combined effects of the WTC gauge coupling plus the 
strong four-fermion coupling, $G_{_{\rm ETC}} \cdot (\bar F F)^2$, arising from spontaneous breaking of ETC to the TC group. 
Here we take $\gamma_m \simeq$ 2, i.e., $(3-\gamma_m) \simeq 1$, for simplicity
for the third-generation SM $f$-fermions like $t, b, \tau$  
which are relevant to the current LHC data.

Thus the TD couplings to $W$ and $Z$ bosons and fermions are 
related to those of the SM Higgs by a simple scaling: 
\begin{eqnarray} 
  \frac{g_{\phi WW/ZZ}}{g_{ h_{\rm SM} WW/ZZ }} 
 = \frac{g_{\phi ff}}{g_{h_{\rm SM} ff}} 
  =\frac{v_{\rm EW}}{F_\phi} 
  \,\, \,
(f=t,b,\tau) . 
\end{eqnarray} 
In addition to the above scaling, 
the couplings to gluon and photon ($G_{\mu\nu}^2$ and $F_{\mu\nu}^2$ terms in ${\cal L}_S$)  
involve the beta functions, $\beta_F(g_s)$ and $\beta_F(e)$, induced from 
$F$-technifermion loops which may be estimated 
at the one-loop level (the same result in the ladder calculations)~\cite{ 
Matsuzaki:2012gd}:   
$   \beta_F(g_s) = \frac{g_s^3}{(4\pi)^2} \frac{4}{3} N_{\rm TC}\,,   
 \beta_F(e) = \frac{e^3}{(4\pi)^2} \frac{16}{9} N_{\rm TC} .
$ 
We thus find the scaling from the SM Higgs for the couplings to 
$gg$ and $\gamma\gamma$, which can approximately be expressed 
 for  
 125 GeV as  
\begin{equation} 
\frac{g_{\phi gg}}{g_{h_{\rm SM} gg}} 
\simeq \frac{v_{\rm EW}}{F_\phi} 
\cdot 
\left( 1 + 2 N_{\rm TC} \right),\qquad
\frac{g_{\phi \gamma\gamma}}{g_{h_{\rm SM} \gamma\gamma}} 
\simeq \frac{v_{\rm EW}}{F_\phi} 
\cdot 
 \left( 1 - \frac{32}{47} N_{\rm TC} \right)  
\,,  
\end{equation} 
where in estimating the SM contributions  
we have incorporated only the top and $W$ boson loops. 

Then we can fit the TD to the LHC data with adjusting the parameters $\frac{v_{\rm EW}}{F_\phi}$ and $N_{\rm TC}$.  The best fit to 
the July 4, 2012 LHC data:
~\cite{Matsuzaki:2012mk} (see also footnote {\it b})
 \begin{eqnarray} 
  \frac{v_{\rm EW}}{F_\phi}\Big|_{\rm best-fit} = 
0.22  
\, (N_{\rm TC}=4), \quad 
0.17  
\, (N_{\rm TC}=5) 
\,. 
\label{best-fit}
  \end{eqnarray}
The salient feature of the TD is that the couplings are smaller than those of the 
SM Higgs by a factor $v_{\rm EW}/F_\phi (\ll 1) $, which is compensated by 
the enhanced gluon fusion production 
due to the colored technifermion contributions mentioned above: $\frac{g_{\phi gg}}{g_{h_{\rm SM}  gg}} 
\gg 1$. This leads to  signal strength compared with that of the SM Higgs in the category of gluon-fusion production:
\begin{eqnarray}
\mu_X = \frac{\sigma_\phi(pp \to \phi) \times {\rm BR}(\phi \to X)}{
\sigma_{h_{\rm SM}}(pp \to h_{\rm SM}) \times {\rm BR}(h_{\rm SM} \to X)} &\simeq& 1 \quad (X=WW/ZZ/\tau^+\tau^-/b
\bar b)\,,
\label{muX}
\\
&\simeq& 1.4  \quad (X=\gamma \gamma) \,,
\label{mu}
\end{eqnarray} 
for a typical value $\frac{v_{\rm EW}}{F_\phi} =0.2$,  where the enhanced diphoton  has extra contributions from
the charged technifermion loop. Eqs.(\ref{muX}, \ref{mu}) are in agreement with the current LHC data. 
Without such extra contributions, we have substantially smaller $\mu_X\ll 1 $ in
 the category of vector-boson fusion/vector-boson associated production, which is consistent with the currently low significance at LHC and should be tested in  future.

Now we come to the evaluation of $F_\phi$, or $\frac{v_{\rm EW}}{F_\phi}$ by specific dynamical calculations.

 In the ladder approximation,  $M_\phi=125$ GeV may be fixed by the value of $F_\phi$ through the PCDC relation, Eq.(\ref{PCDC}),
with the ladder criticality condition $N_{\rm TF}=4N_{\rm TC}$~\cite{Appelquist:1996dq}, where $m_F$ is related to the decay constant of the technipion $F_\pi$ $ (=\frac{v_{\rm EW}}{\sqrt{N_D}},
\,N_D=4$ for 1FM),  through the Pagels-Stokar formula:
 $
  F_\pi^2 = \kappa_F^2 \frac{N_{\rm TC}}{4 \pi^2} m_F^2\,, 
$
with the ladder value $\kappa_F \simeq 1.4$~\cite{Hashimoto:2010nw}. 
Then we estimate up to the 30 \% uncertainties of the ladder 
approximations
 $ 
 \frac{v_{\rm EW}}{F_\phi}
\simeq (0.1 - 0.3)  
\cdot \left( \frac{N_D}{4} \right) \left( \frac{M_\phi}{125\,{\rm GeV}} \right) ,
$ 
which  
is  consistent with Eq. (\ref{best-fit}).

Another method to estimate $F_\phi$ is a (bottom up) holographic computation~\cite{Haba:2010hu} which is a deformation of the successful holographic QCD~\cite{DaRold:2005zs
}, 
with the anomalous dimension $\gamma_m =0$ replaced by $\gamma_m=1$ via the bulk mass term of the bulk scalar. In addition we also introduce a bulk gluon field  to match the current correlators to 
those of the Operator Product Expansion in QCD ($\gamma_m=0$) and WTC ($\gamma_m=1$). The model is very successful in reproducing the QCD phenomenology~\cite{Haba:2010hu,Matsuzaki:2012xx}.  The inclusion of the fully {\it nonperturbative gluonic dynamics} via gluon condensate  is a distinct feature compared with the ladder approximation which totally ignores non-ladder dynamics most notably the full gluonic dynamics.

It was shown in WTC~\cite{Matsuzaki:2012xx} that the massless limit of TD does exist when the technigluon condensate becomes large,
\begin{equation}
\frac{M_\phi}{4\pi F_\pi} 
\simeq  \sqrt{ \frac{3}{N_{\rm TC}}} \frac{\sqrt{3}/2}{1+ G} \rightarrow 0
\qquad 
\left(G \sim \frac{\langle \alpha G_{\mu\nu}^2   \rangle}{F_\pi^4}\rightarrow \infty \right)
\,,
\end{equation} 
which can be identified with the limit of conformality $\beta(\alpha) \rightarrow 0$. Actually, the mass $M_\phi=125$ GeV corresponds to $G \simeq 10$,
which is compared with the QCD best fit $G_{\rm QCD} \simeq 0.25$~\cite{Haba:2010hu,Matsuzaki:2012xx}.
Hence the light TD can naturally be realized in the holographic method in contrast to the ladder estimate of the PCDC relation where the TD gets decoupled in the massless limit~\cite{Haba:2010hu,Hashimoto:2010nw}. 
Also note that $M_\phi$ is estimated fairly independently of the value of $F_\phi$ which is also contrasted to the ladder PCDC estimate.

Most amazingly, the model predicts:~\cite{Matsuzaki:2012xx} 
\begin{equation} 
 \frac{F_\phi}{F_\pi} \simeq (3-\gamma_m) \sqrt{\frac{N_{\rm TF}}{2}} 
 \,,  
 \end{equation} independently of the holographic parameters in the light TD mass region, $M_\phi/(4\pi F_\pi) \ll 1$, in a way fairly insensitive to the precise value of the mass of $M_\phi$.
 Then we have 
 $\frac{v_{\rm EW}}{F_\phi} = \frac{1}{2} \sqrt{2N_D/N_{\rm TF}}
 $ for  $\gamma_m=1$.

Another nice feature of this
holography is~\cite{Matsuzaki:2012xx}  that we can tune the S parameter, say $S<0.1$, fairly independently of the TD phenomenology for the 125 GeV Higgs data at LHC.

 Incorporating typical $\sim $30\% corrections into the holography 
coming from the next-to-leading order terms in 
$1/N_{\rm TC}$ expansion,  
we may estimate the TD decay constant $F_\phi$ to get~\cite{Matsuzaki:2012xx} 
$ 
\frac{v_{\rm EW}}{F_\phi}  
\simeq 0.2 -0.4 , 
$ 
which is consistent with the best-fit to the LHC data Eq.(\ref{best-fit}). Note also
that  in spite of the qualitative difference in the massless limit, the holographic result numerically coincides with the ladder estimate mentioned above, 
for the particular value $M_\phi \simeq 125$ GeV. 

  \section{Discovering Walking Technicolor on the Lattice}
  \label{sec:Lattice}
  We have seen that the ladder and the holography describe the technidilaton consistently with the 125 GeV Higgs. However, both computations suffer from large uncertainties.
    Since the WTC is a strongly coupled gauge theory, the most reliable method  will be the lattice simulations.  
  
 We actually started two years ago such lattice simulations at KMI 
 (``LatKMI Collaboration'')  in search for the WTC.
 On behalf of the LatKMI Collaboration
 I here present our lattice results~\cite{Aoki:2012eq,Aoki:2013pca,Aoki:2013xza} on the so-called large $N_f$ QCD~\cite{Iwasaki:1991mr}, a class of $SU(3)$ gauge theories with $N_f$ fermions having a degenerate mass $m_f$ 
 which is eventually extrapolated to the chiral limit. 
 
 These theories are expected to become a walking theory at a certain large $N_f$ less than $N_f=16.5$ where the asymptotic freedom is lost. The Caswell-Banks-Zaks infrared fixed point at two loop appears at $N_f \simeq 8$, which may be washed out by the dynamical generation of the fermion mass
 as was suggested by the ladder analysis for  $N_f < N_f^{\rm cr} \simeq 4 N_c =12$~\cite{Appelquist:1996dq} 
.  Thus we expect that the walking theory may be realized in between $N_f=8$ and $N_f=12$.
   Of particular interest is to find a candidate for such a walking theory and, 
 most urgently, a light composite flavor-singlet scalar as a candidate for the technidilaton.  
 
 We have studied $N_f=4,8,12,16$ on the same setup to make a systematic study of 
 the spectrum throughout different $N_f$'s. Our simulations are based on the tree-level Symanzik gauge action and the highly improved staggered quarks (HISQ) action
 which, as expected, was actually observed in our simulations to have a good flavor (taste) symmetry and small discretizing errors in the staggered fermions. We here present selected results for
 $N_f=4,8,12$ where the simulation parameters (in units of lattice spacing $a$), bare coupling $\beta=6/g^2$, lattice size $L\times T$ and the bare fermion mass $m_f$,   
 are as follows:\\ 
$N_f=4$~\cite{Aoki:2013xza}: 
\hspace{0.46cm} $\beta =3.7$, $L^3\times T = 12^3\times 18 - 20^3\times 30$, $m_f=0.005-0.05$, \\
$N_f=8$~\cite{Aoki:2013xza}:
\hspace{0.47cm} $\beta=3.8$, $L^3\times T = 18^3\times 24 - 36^3\times 48$, $m_f=0.015-0.16$,  \\
$N_f=12$~\cite{Aoki:2012eq,Aoki:2013pca}:
$\beta=3.8$, $L^3\times T = 18^3\times 24 - 36^3\times 48$, $m_f=0.04-0.2$ , \\
\hspace {2.63cm}$\beta=4.0$, $L^3\times T = 18^3\times 24 - 36^3\times 48$, $m_f=0.05-0.2$,\\
with $L$ being large enough for each $m_f$ such that $m_f \gg 1/L$ and $m_\pi L > 6$ ($L > 4$ for $N_f=4$).

The gross feature of each $N_f$ can be seen from typical data on the ratio $F_\pi/M_\pi$ of the decay constant $F_\pi$ and the
mass $M_\pi$ of $\pi$ (corresponding to the Nambu-Goldstone pion in QCD) given in Fig.\ref{compare}:
 \begin{figure}[h]  
\includegraphics[width=5.0cm]{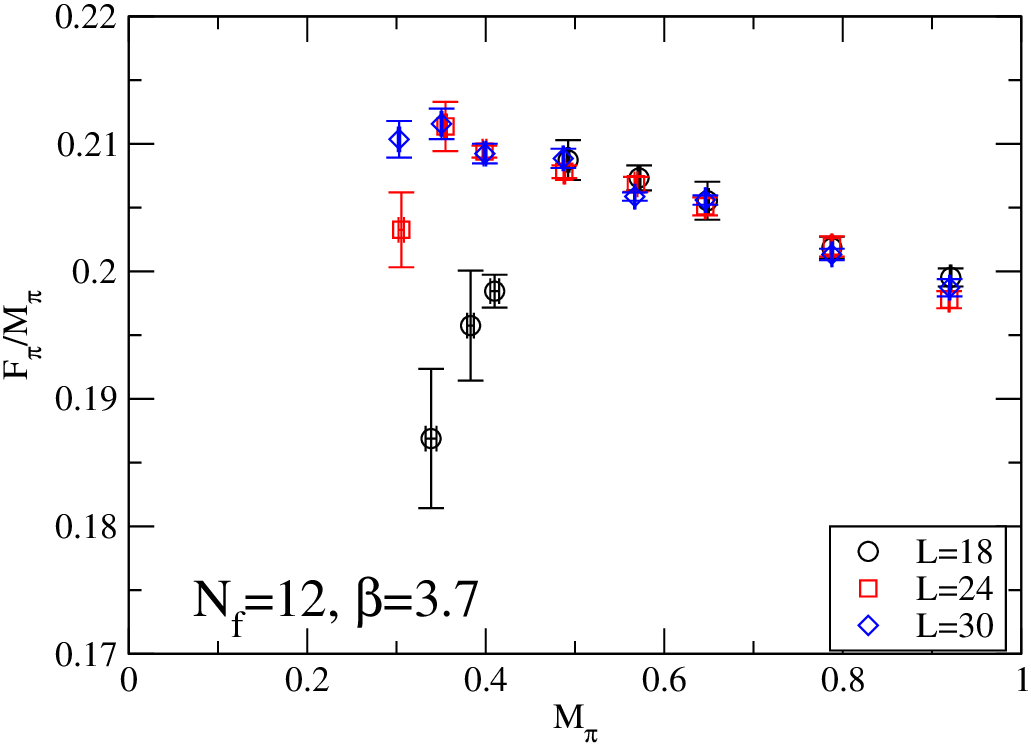}
\includegraphics[width=5.0cm]{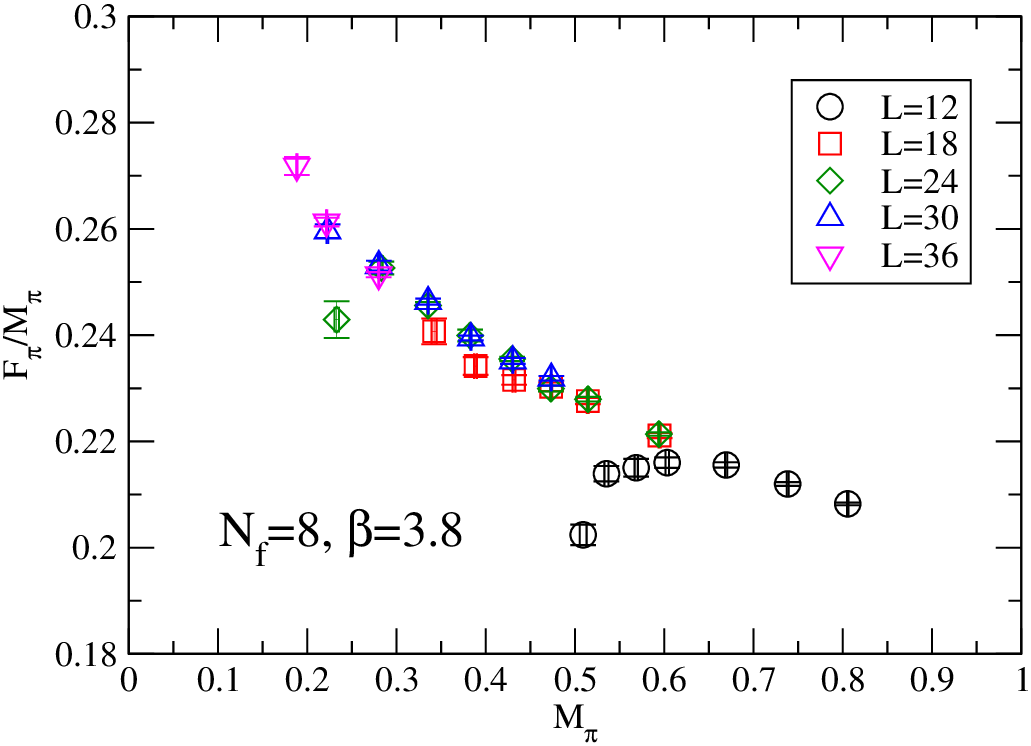}
\includegraphics[width=5.0cm]{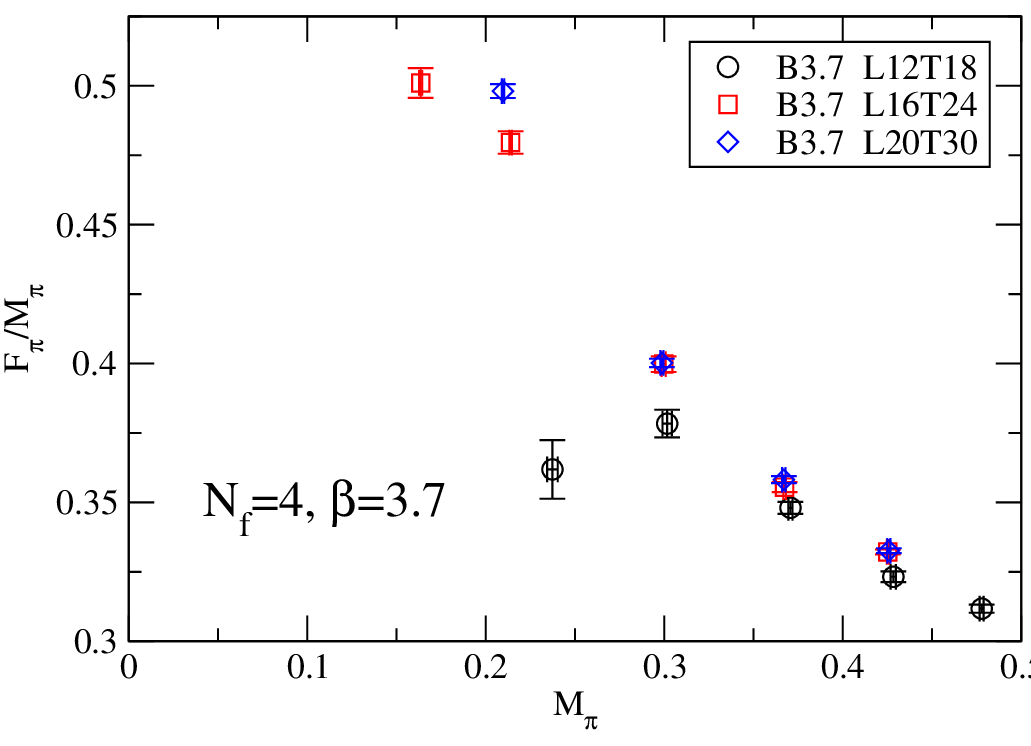}
\caption{Typical chiral behavior of $N_f=4,8,12$ seen by $F_\pi/M_\pi$ when approaching $M_\pi \rightarrow 0$. }
\label{compare}
\end{figure}     
For $N_f=12$ (left panel),  $F_\pi/M_\pi$ stays non-zero 
constant value near the chiral limit, which is consistent 
with the hyperscaling relation in the conformal window without spontaneous chiral symmetry brewing, while  it diverges there for $N_f=8$ (center panel) and $N_f=4$ (right panel), implying the chiral symmetry broken phase. 
Actually we made more detailed analysis for $N_f=12$, which is consistent with the finite-size hyperscaling relation in the conformal window,  
having the anomalous dimension $\gamma_m \sim 0.4-0.5$ somewhat smaller  than that expected for the walking theory~\cite{Aoki:2012eq}. 
We also confirmed 
chiral symmetry is spontaneously broken for $N_f=4$
~\cite{Aoki:2013xza}.

In contrast,  the $N_f=8$ data show a dual picture~\cite{Aoki:2013xza}:  They are consistent  with chiral perturbation theory for smaller mass region
indicating clear signals that 
\begin{equation}
M_\pi=0,\quad F_\pi \ne 0, \quad
M_\rho \ne 0, \quad
<\bar q q > \ne 0 \qquad (m_f=0.015 - 0.04)\,,
\end{equation}
in the chiral limit extrapolation,
while the data for intermediate mass region 
are all consistent with the hyperscaling relation 
\begin{equation}
L M_H= f_H (L m_f^{\frac{1}{1+\gamma}}) \quad (m_f=0.05-0.16)\,,
 \end{equation}  
  with $\gamma({M_\pi})  =0.57$, 
$\gamma_m (F_\pi) \simeq 0.93$, $\gamma_m (M_\rho) \simeq 0.80$ , which should be a remnant of the conformality
even if the chiral symmetry is spontaneously broken.
\footnote{
The hyperscaling relation away from the chiral limit should have $m_f$ corrections
even in the conformal window  as was demonstrated by the ladder SD equation.~\cite{Aoki:2012ve}
After such corrections we have a universal
 $\gamma=0.62 - 0.97$, with the value depending on the form of the corrections.~\cite{Aoki:2013xza} 
 }  
This can be understood naturally if the $N_f=8$ theory is a walking theory described by the picture
in Fig.\ref{alpha-beta}: Our lattice simulations introduce the explicit mass $m_f$ as a probe, which may be either in the region III or region II of the picture (left panel), depending on 
the cases $m_f \ll m_F$ or $m_F \ll m_f \ll \Lambda_{\rm TC}$, respectively, where $\Lambda_{\rm TC}$ and $m_F$ in this case read respectively the intrinsic scale $\Lambda_{\rm QCD}$ and a typical scale of the spontaneous chiral symmetry breaking, roughly corresponding to the above boundary value $m_f\simeq 0.04-0.05$. Such a dual feature is in fact 
 a desired feature of the walking technicolor and $N_f$ QCD would be a dynamics as a candidate for the WTC.

\begin{figure}[h]  
\includegraphics[width=5.0cm]{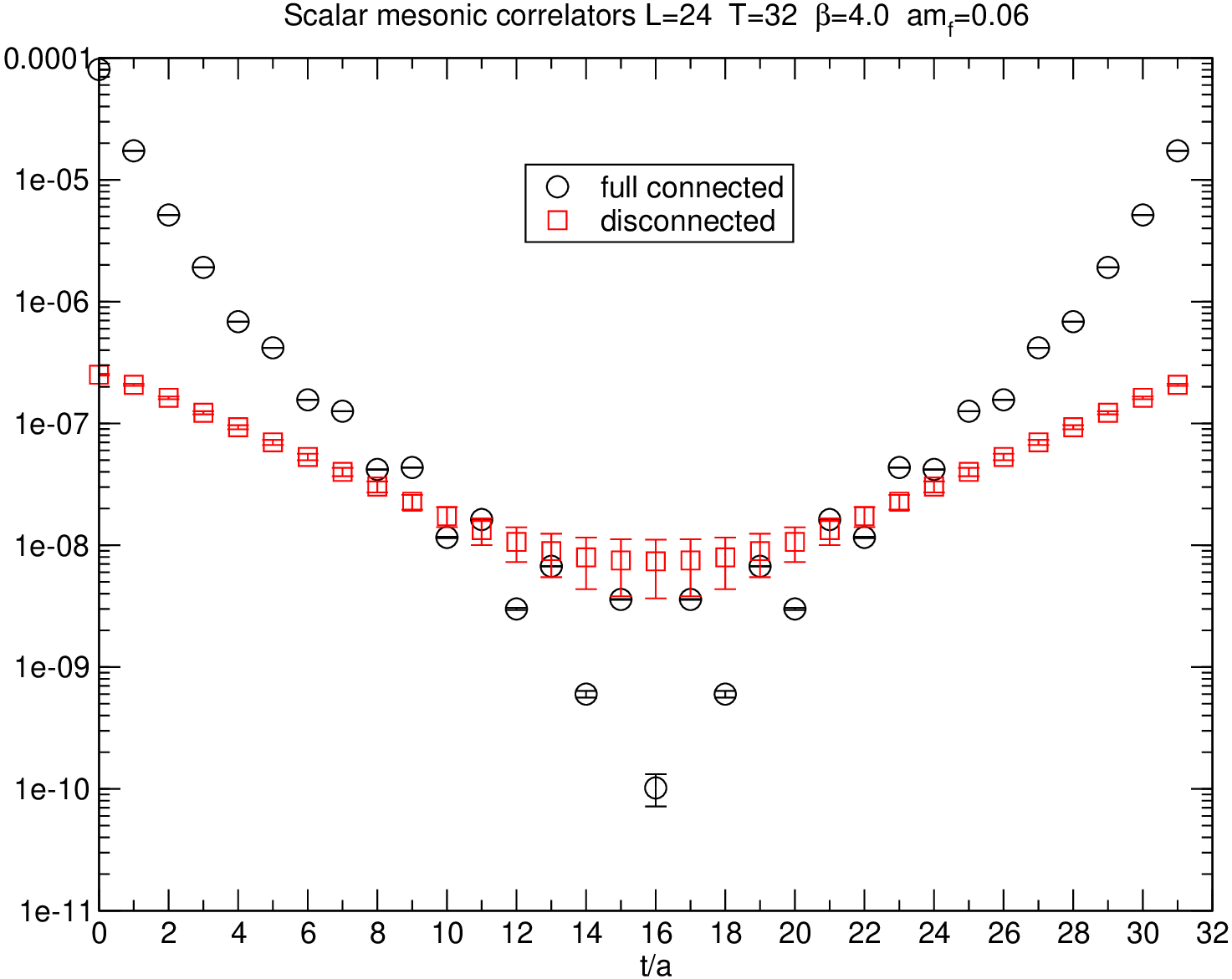}
\includegraphics[width=5.0cm]{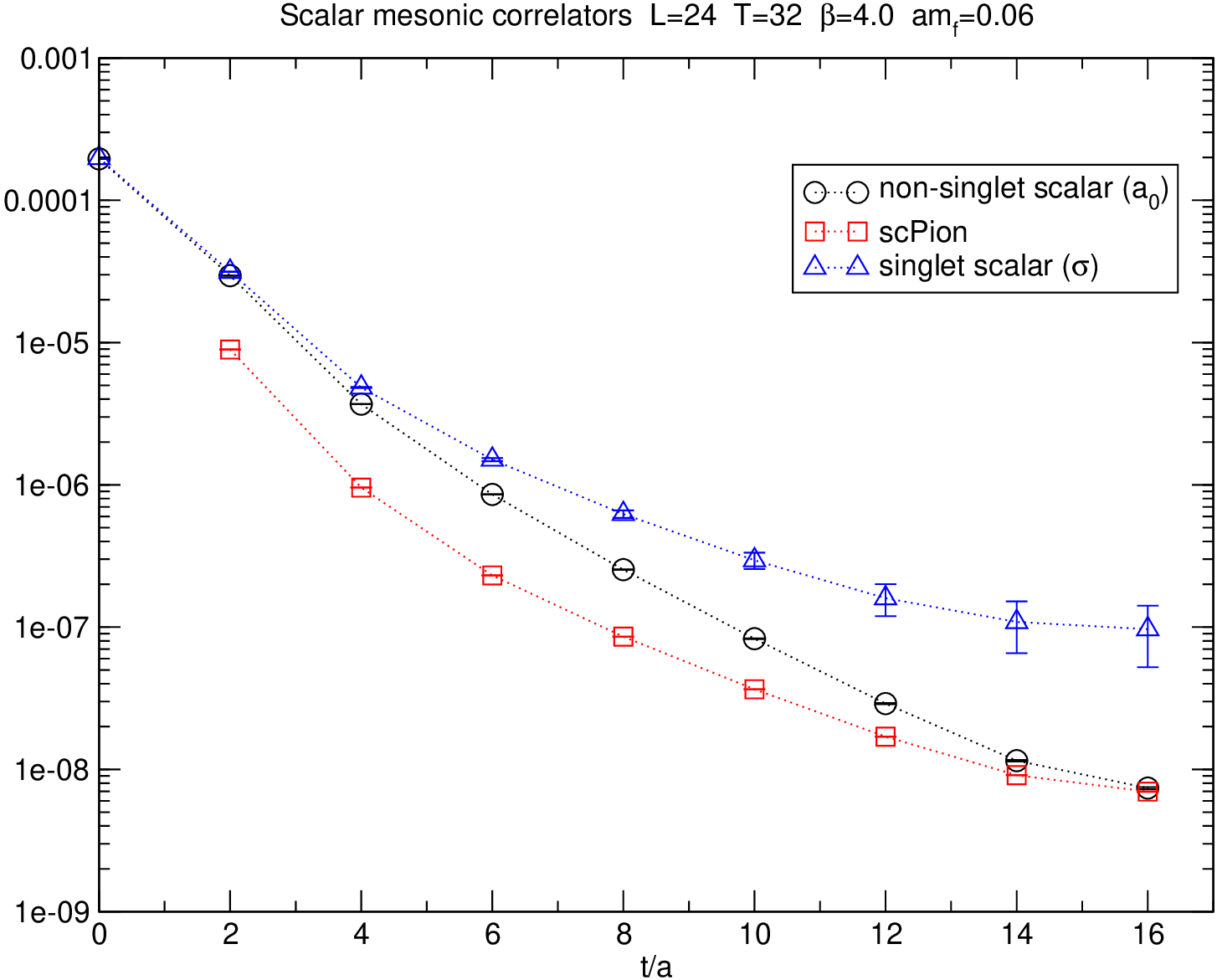}
\includegraphics[width=5.0cm]{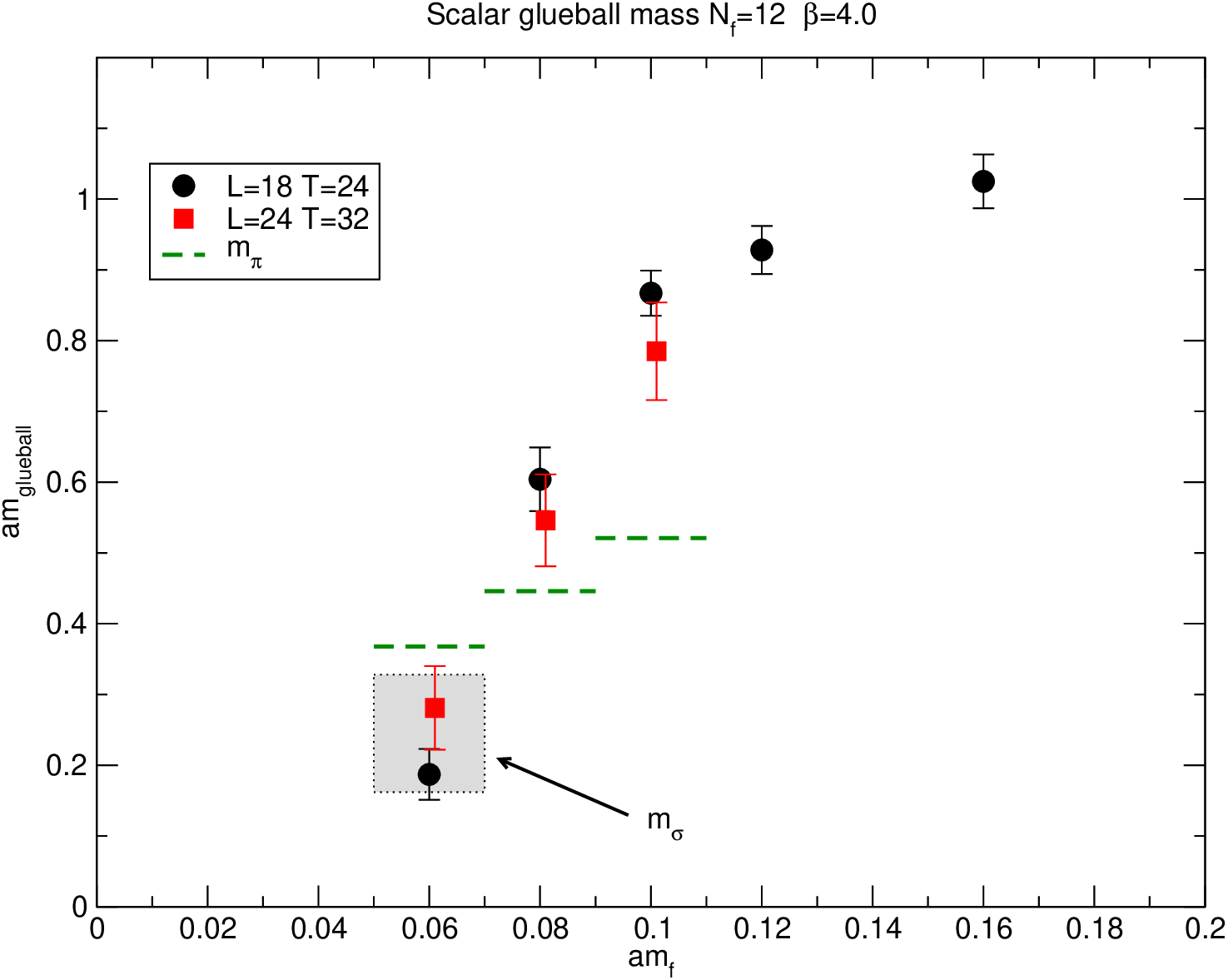}
\caption{A clean signal of $M_\sigma$
 observed both in the ferminic and gluonic operators at $m_f=0.06$ for $N_f=12$.  
 (Left): Full connected and disconnected correlators.  (Center): Correlators for different channels from the full connected correlator,
 where the $\sigma$ channel was extracted from a disconnected correlator. The mass $M_\sigma$ is clearly lower than the non-singlet scalar $a_0$.
 (Right): $M_\sigma$  lighter than $M_\pi$  (indicated by dashed green line) is compared also with the glueball mass (heavier than $M_\pi$) at $m_f$ other than $m_f=0.06$. }
\label{scalar}
\end{figure}

Now I come to a scalar composite lighter than $\pi$  in $N_f=12$.~\cite{Aoki:2013pca}
In Fig.\ref{scalar}  we present the results of a flavor-singlet scalar meson $\sigma$ ($\bar q q$ bound state)  
from the fermionic operators, both connected and disconnected correlators. We found a  clean signal for the mass $M_\sigma$ lighter than $M_\pi$F$M_\sigma < M_\pi$,  in the
full flavor-singlet correlator for $m_f=0.06$, which was also observed in the gluonic correlators. Such a scalar lighter than
$\pi$ should be reflecting the conformality of the dynamics in the energy region $m_F={\cal O}(F_\pi)<\mu<\Lambda_{\rm QCD}$.
Although we observed that $N_f=12$ is
consistent with the conformal window without chiral symmetry breaking and hence is not a candidate for the WTC as
a model of electroweak symmetry breaking, existence of such a light scalar for $N_f=12$ would also imply a similar light scalar, 
 to be identified with the composite Higgs, or technidilaton, in $N_f=8$, which has a hyperscaling relation, a remnant of conformality similar to $N_f=12$, for the intermediate
$m_f$ region as mentioned above. Work is in progress for $N_f=8$.

In conclusion, we have discussed technidilaton as a salient feature of the walking technicolor to be tested at 
LHC and on the lattice. Ladder and holography give similar results both consistent with the current data of the 125 GeV Higgs at LHC.
We have discussed that $N_f=8$ QCD is a walking theory having a large anomalous dimension near unity. We
observed a light composite scalar on the lattice at $N_f=12$, which may suggest existence of a similar light scalar at $N_f=8$ whose
dynamics above the chiral symmetry breaking scale is similar to $N_f=12$. Work is in progress towards this point. We hope to have
more precise lattice data for WTC to be tested by the up-graded LHC results coming two years later. We will see.

\section*{Acknowledgments}
 I would like to thank S. Matsuzaki and LatKMI Collaboration members for collaborations on the recent results presented in this paper.
  This work is supported in part by the JSPS Grant-in-Aid for Scientific Research (S) No.22224003 and 
 (C) No.23540300.

\end{document}